\documentclass[a4paper,11pt]{article}
\usepackage{pos}
\usepackage{float}
\usepackage{wrapfig}

\usepackage{lineno}

\usepackage{enumitem}

\title{Study of solar activity with AERA at the Pierre Auger Observatory}

\author*[a]{R. M. de Almeida}

\affiliation[a]{Instituto de Física, Universidade Federal do Rio de Janeiro,\\
  Rio de Janeiro, Caixa postal 68528, Rio de Janeiro, Brazil}

\onbehalf{for the Pierre Auger Collaboration$^b$}

\affiliation[b]{Observatorio Pierre Auger, Av.\ San Mart{\'\i}n Norte 304, 5613 Malarg\"ue, Argentina\\
Full author list: {\rm\url{https://www.auger.org/archive/authors_icrc_2025.html}}}

\emailAdd{spokespersons@auger.org}

\abstract{Solar activity events release vast amounts of energy, including radio waves, X-rays, ultraviolet radiation, and energetic particles, which interact with the ionosphere of the Earth and can disrupt radio wave propagation, affecting radio communications. They can either enhance reflections, improving long-distance terrestrial communications, or cause signal degradation and absorption, respectively, depending on whether the increased ionization affects the upper or lower layers of the ionosphere. In the first case, the solar cycle modulates the Maximum Usable Frequency (MUF), the highest frequency usable for radio communication between two Earth-based points. The Auger Engineering Radio Array (AERA) of the Pierre Auger Observatory was developed to measure the radio emission from extensive air showers in the $30-80$ MHz band. We examine the impact of solar activity on AERA data collected over approximately 11 years. We report the detection of different types of solar radio bursts and we investigate how increased solar radiation - particularly in the X-ray and extreme ultraviolet bands - also affects measurements in the AERA energy band. Our results show a remarkable correlation between the MUF and the broadband noise observed in the $30-40$ MHz frequency range. Radio blackouts are also observed in AERA spectrograms in coincidence with those reported by the National Oceanic and Atmospheric Administration (NOAA). Additionally, we performed a search for temporal coincidences between AERA data and independent observations of solar radio burst events from the e-CALLISTO network and the SWAVES instrument. These findings highlight the complex interplay between solar activity and radio wave propagation, which is also relevant for cosmic-ray detection.}

\ConferenceLogo{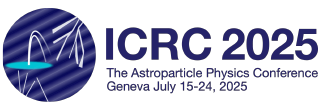}

\FullConference{39th International Cosmic Ray Conference (ICRC2025)\\
 15–24 July 2025\\
Geneva, Switzerland\\}

\begin{document}
\maketitle

\section{Introduction}
\label{sec:intro} 

The Auger Engineering Radio Array (AERA)~\cite{AERA} is a facility of the Pierre Auger Observatory, located in Malargüe, Argentina, dedicated to detecting radio emissions from extensive air showers induced by high-energy cosmic rays. AERA comprises a network of 153 autonomous stations spread over an area of approximately 17 km$^2$, operating in the frequency range of 30 to 80 MHz. Although its primary purpose is focused on cosmic-ray detection, the periodic measurements collected by the antennas can also be utilized to investigate phenomena associated with solar activity.

The Sun has an approximately 11-year activity cycle, characterized by alternating periods of low and high intensity, marked by variations in the number of sunspots, solar flares, coronal mass ejections (CMEs), and changes in solar irradiance. During the maximum of this cycle, the significant increase in ultraviolet and X-ray radiation intensifies the ionization process in the Earth’s ionosphere, especially in the upper layers. This phenomenon leads to an increase in electron density, raising the Maximum Usable Frequency (MUF), which represents the highest frequency that can be used for radio communication between two points on Earth, taking ionospheric conditions into account. When the electron density is high, the MUF is also high, allowing higher frequencies to be used for long-distance radio communication as these waves are reflected by the ionosphere. For frequencies above the MUF, the atmosphere becomes transparent, allowing the signal to pass through the ionosphere.

In addition to its effects on the MUF, periods of intense solar activity often trigger solar flares, characterized by the sudden and intense release of electromagnetic radiation. When this radiation reaches Earth, it penetrates deeply into the atmosphere, significantly impacting the internal layers of the ionosphere, particularly the D layer. Under typical conditions, this layer is not involved in the reflection of radio waves. However, during intense ionization induced by solar flares, it becomes a critical absorption region. Radio waves propagating through this ionized medium experience energy loss due to enhanced collision rates with free electrons, resulting in significant signal attenuation. This process can culminate in a phenomenon known as a radio blackout, where high-frequency radio communications, primarily within the 3 to 30 MHz range, are partially or completely disrupted.

Apart from radio blackouts, the intense release of energy during a flare can accelerate charged particles along the solar magnetic field lines, triggering radio wave emissions over a wide range of frequencies, a phenomenon known as Solar Radio Bursts (SRBs). Although solar flares are a prerequisite for the occurrence of SRBs, not all flares produce them. These events are classified into five main categories: Types I, II, III, IV, and V ~\cite{TypeI,TypeII,TypeIII,TypeIV,TypeV,KUNDU}, based on their spectral characteristics. Each type exhibits distinct signatures in terms of frequency drift, duration, and association with specific solar phenomena, such as coronal mass ejections (CMEs) and solar flares.

In this work, we analyze approximately 11 years of AERA data, encompassing two solar maximum periods: the first during the maximum of Solar Cycle 24 in 2014, and the second during the maximum of Solar Cycle 25 occurring around $2024-2025$, to investigate how solar activity influences the recorded radio signals. We use data from the Butterfly antennas ~\cite{AERA}, commonly referred to as bow-tie models, consisting of two triangular arms oriented along the East-West and North-South relative to magnetic North. Our results reveal a significant increase in broadband noise within the $30-40$ MHz range, correlated with the rise of the MUF during periods of high solar activity. We also search for radio blackouts coinciding with those reported by the National Oceanic and Atmospheric Administration (NOAA), as well as solar radio bursts observed in correlation with independent measurements from the e-CALLISTO network and the SWAVES instrument aboard the STEREO-A satellite.

The article is organized as follows: Section \ref{section-muf} presents the correlation observed between the MUF and the increase in noise levels detected by the AERA antennas. Section \ref{section-SRB} describes our search for radio blackouts associated with events reported by NOAA. Section \ref{section-Algorithm-validation} describes the identification of solar radio bursts in AERA data through temporal coincidence with independent observations from e-CALLISTO and SWAVES. Finally, conclusions are presented in Section \ref{conclusions}.

\section{Correlation between MUF and broadband noise measured by AERA within the \textbf{30\,--\,40 MHz} MHz frequency range}
\label{section-muf}

The increase in the MUF triggered by intense solar activity facilitates the reflection of higher-frequency radio waves that would normally pass through the ionosphere without reflection. This phenomenon causes interference in ground-based systems sensitive to this frequency range. Normally, the MUF in Malargüe does not exceed 30 MHz, which is the lower limit of AERA’s frequency range. However, during periods of high solar activity, the MUF can surpass this threshold, allowing radio waves from distant terrestrial sources to be reflected by the ionosphere and impinge on the Auger Observatory.

In this context, we investigated how MUF variations affect AERA measurements, particularly in the lower operational range ($30-40$ MHz), over a decade of data collection, from 2014 to 2024. For this analysis, we employed the International Reference Ionosphere (IRI) model~\cite{IRI}, which utilizes the Maximum Usable Frequency monitored locally by various stations around the world and interpolates this data to estimate the MUF in any given region\footnote{In this study, we utilize the MUF in the Malargüe region with coordinates: Latitude: $-35.47^{\circ}$, Longitude: $-69.58^{\circ}$.}. We observed a clear and significant correlation between periods of high MUF and substantial increase in noise levels recorded by AERA antennas.

\begin{figure}[H]
    \vspace{-0.1cm}
        \centering
                \includegraphics[scale=0.30]{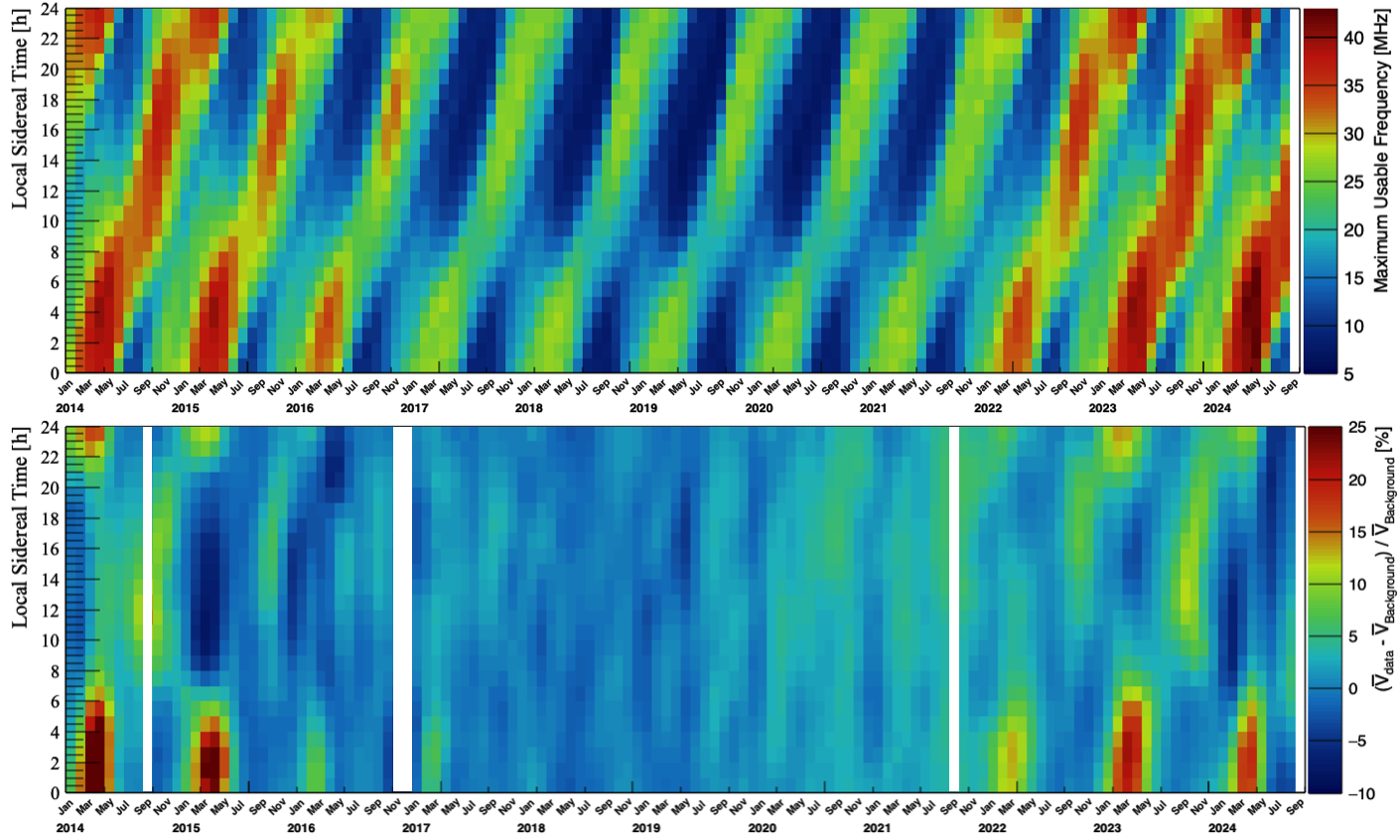}
         \caption{\footnotesize{Comparison between the Maximum Usable Frequency (MUF) and the average percentual difference in signal amplitude measured by AERA stations.
The top panel shows the monthly MUF recorded in the Malargüe region from 2014 to Semptember 2024 as a function of LST. The bottom panel displays the average percentual difference in the 30–40 MHz signal amplitude relative to the background level, defined as the average signal measured during 2018 (solar minimum) for each hour in LST. This difference represents the combined effect from both east-west and north-south channels. White bands indicate periods with no available data. A clear correlation between MUF variations and the observed differences in signal amplitude is evident.}}
\label{Fig:mufxtime}
\vspace{-0.3cm}
 \end{figure}

To illustrate this correlation, Figure \ref{Fig:mufxtime} presents the comparison between the MUF and the average percentual difference in signal amplitude measured by AERA stations. The top panel shows the monthly MUF recorded in the Malargüe region from 2014 to September 2024 as a function of Local Sidereal Time (LST). The bottom panel displays the monthly average percentual difference in the $30-40$ MHz signal amplitude $\bar{V}_{\rm{data}}(t)$, calculated from the combined measurements of all Butterfly stations across both east-west and north-south channels. This difference is referenced against the background level $\bar{V}_{\rm{background}}(t)$, defined as the average signal measured during 2018 (solar minimum) for each hour in LST. The white bands represent periods where no data is available for the entire LST range during that month. The figure shows that high MUF values are observed only during periods of increased solar activity. During these same periods, an increase in the average signal amplitude measured by AERA is also evident. This observation strongly suggests that the detected noise is caused by enhanced solar-induced ionization in the upper ionospheric layers, which facilitates the reflection of radio waves from distant terrestrial sources back to the surface, thereby affecting the signals recorded by AERA.

\section{Search for radio blackouts in temporal coincidence with NOAA reports} 
\label{section-SRB}

As discussed in Section \ref{sec:intro}, solar flares are known to release intense bursts of electromagnetic radiation, which, when directed towards Earth, can significantly modify the radioelectric environment and affect the propagation of radio waves. The intense ionization induced in the Earth’s ionosphere, particularly in its lower layers, can lead to temporary disruptions in terrestrial communications, a phenomenon known as radio blackouts. These blackouts primarily affect the $3-30$ MHz frequency band. In AERA’s operational range of $30-80$ MHz, however, such disruptions are less common, as only exceptionally strong solar flares are capable of producing noticeable effects. Solar flares are classified according to their X-ray intensity into four main categories: X, representing the highest intensity, M for moderate events, C, and B, which correspond to progressively lower intensities \cite{GOES_NOAA}. In turn, radio blackouts are primarily associated with Class X and M solar flares and are categorized by NOAA on a scale from R1 to R5, indicating levels of severity from minor to extreme \cite{NOAA_scale}.

\begin{figure}[H]
     \vspace{-0.2cm}
        \centering
                \includegraphics[scale=0.40]{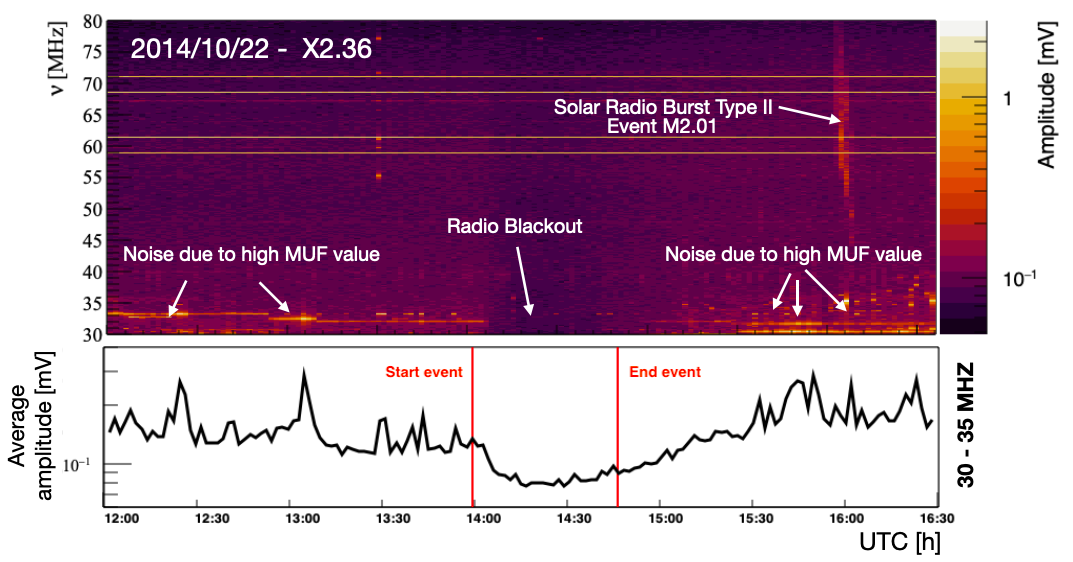}\quad
         \caption{\footnotesize{Average dynamic frequency spectrum as a function of UTC, captured by all Butterfly antennas around the time of a particular event which occurred in October 2014 classified as X2.36. The spectrogram illustrates a radio blackout occurrence during the event, along with additional noise attributed to the high MUF value. The bottom panel displays the signal average across all frequencies from 30 to 35 MHz. Vertical red lines indicate the start and end of the solar flare as reported by NOAA.}}
         \label{Fig:event_n3}
         \vspace{-0.3cm}
 \end{figure}

In this work, we investigated 16 solar flare events reported by NOAA between 2014 and 2024 that occurred during daylight hours in South America. According to NOAA, these events were classified as R2 and R3 levels of radio blackouts based on its scale, causing substantial disruptions in radio wave propagation and temporary interruptions in signal transmission. In our analysis, we specifically searched for signatures of radio blackouts in AERA data. Although radio blackouts primarily affect frequencies below 30 MHz, out of the 16 events reported by NOAA, we observed radio blackouts in six cases within the AERA data. Interestingly, we also observed 11 solar radio bursts in temporal correlation with these NOAA-reported events. As an illustrative example, we highlight a particular event detected which occurred in October 2014 and led to a radio blackout observable in the AERA data, as depicted in Figure \ref{Fig:event_n3}. The impact of this event is evident through a reduction in the amplitude of the AERA signals precisely during the period of the flare, as shown in both panels of the figure. Moreover, this solar flare triggered a radio blackout in telecommunications frequencies below 30 MHz, as reported by NOAA. The dynamic spectra also reveal significant amplitude increases around 30 MHz, linked to elevated MUF values, a typical signature of the heightened solar activity during 2014. The bottom panels present the average signal integrated over the $30-35$ MHz band highlighting the pronounced effects of the radio blackout. Interestingly, a solar radio burst associated with an M2.01 solar flare, marked in the Figure, was also observed in AERA data. A more detailed analysis of SRBs is presented in the next section.

\section{Search for SRBs in temporal coincidence with e-CALLISTO and SWAVES} 
\label{section-Algorithm-validation}

As discussed in Section \ref{sec:intro}, SRBs are phenomena that occur when charged particles are accelerated along the solar magnetic field lines, generating radio wave emissions over a broad range of frequencies. If these signals fall within AERA’s frequency range, they can produce distinctive signatures in the recorded data. However, detecting these events requires that South America, where the Pierre Auger Observatory is located, be illuminated by sunlight at the time of the flare, as the emitted radiation reaches Earth in about eight minutes. Inspired by the study reported in~\cite{RNO-G}, we performed a search for temporal coincidence between SRBs detected by AERA and those recorded by the e-CALLISTO network during the period from 2021 to 2023 and SWAVES instrument aboard NASA’s STEREO-A satellite during the period from 2015 to 2023.

The global e-CALLISTO network~\cite{Callisto} is an international array of radio spectrometers specifically designed for real-time monitoring of solar radio emissions. Comprising antennas distributed across multiple countries, the network continuously observes solar activity over a broad spectrum of frequencies, enabling the detection of SRBs and other transient solar phenomena.

In addition to the terrestrial e-CALLISTO antennas, the SWAVES instrument aboard NASA’s STEREO-A satellite~\cite{Swaves}, is an interplanetary radio burst tracker that traces the generation and evolution of traveling radio disturbances from the Sun to the orbit of Earth. Operating in the low-frequency range of 1 MHz to 16 MHz, SWAVES captures independent measurements of solar radio bursts with high sensitivity.

To identify SRBs in AERA data, we developed an automatic detection algorithm that operates on the daily averaged spectrograms recorded by all antennas.  Given AERA’s periodic data acquisition, with measurements taken every 100 seconds (defining the time resolution of the spectrogram), solar radio bursts appear in the spectrograms as localized increases in signal intensity over only a few time bins. Thus, an SRB appear as sharp, well-defined peaks within the spectrogram, standing out distinctly from the background noise. The first step of the method to detect SRBs in AERA data consists of calculating the difference of the signal intensity for each consecutive frequency bin $\nu$ and time bin $t$, as defined by the expression:
\begin{equation}
    \Delta S = S(\nu, t) - S(\nu, t-1),
\end{equation} where $S(\nu, t)$ represents the measured signal at a specific frequency bin $\nu$ and time bin $t$, where each frequency bin corresponds to a resolution of approximately 0.175 MHz and each time bin to an interval of 100 seconds. This calculation captures the variations in signal intensity across both time and frequency dimensions. This procedure is applied to each daily spectrogram from 2014 to 2024, generating a distribution of these differences for every frequency bin throughout the entire period analyzed. To isolate potential SRB signals from background noise, only values of $\Delta S$ above the 99th percentile ($\Delta S^{99\%}$) of this distribution are selected as candidate events.  The algorithm then searches for SRB candidates by applying two additional conditions. A signal is considered a potential SRB if the range of consecutive frequency bins with $\Delta S$ above $\Delta S^{99\%}$\ exceeds 15 MHz or if the number of frequency bins with $\Delta S$ above  $\Delta S^{99\%}$ corresponds to a range of at least 30 MHz. If either condition is satisfied, the event is stored as a potential SRB for further analysis. To investigate whether these candidate events coincide with solar radio bursts observed by other instruments, we performed a temporal coincidence analysis. In this context, a temporal coincidence is defined when a solar radio burst detected by the e-CALLISTO network or the SWAVES instrument is also identified in AERA data within the same time window, including the respective timing uncertainties associated with the temporal resolution of each instrument. For the analysis of temporal coincidences with the e-CALLISTO network, antennas operating in frequency bands close to AERA’s range were selected and only SRBs observed simultaneously by at least two antennas located in different countries were considered. This selection criterion was applied to reduce geographical biases and increase the reliability of the data. Additionally, the analysis was restricted to events occurring during daylight hours at the AERA site in Malargüe. Out of the 639 SRBs detected by e-CALLISTO that satisfied these requirements, we identified 204 temporal coincidences with events automatically detected by

\begin{figure}[H]
        \centering
        \vspace{-0.1cm}
                \includegraphics[scale=0.25]{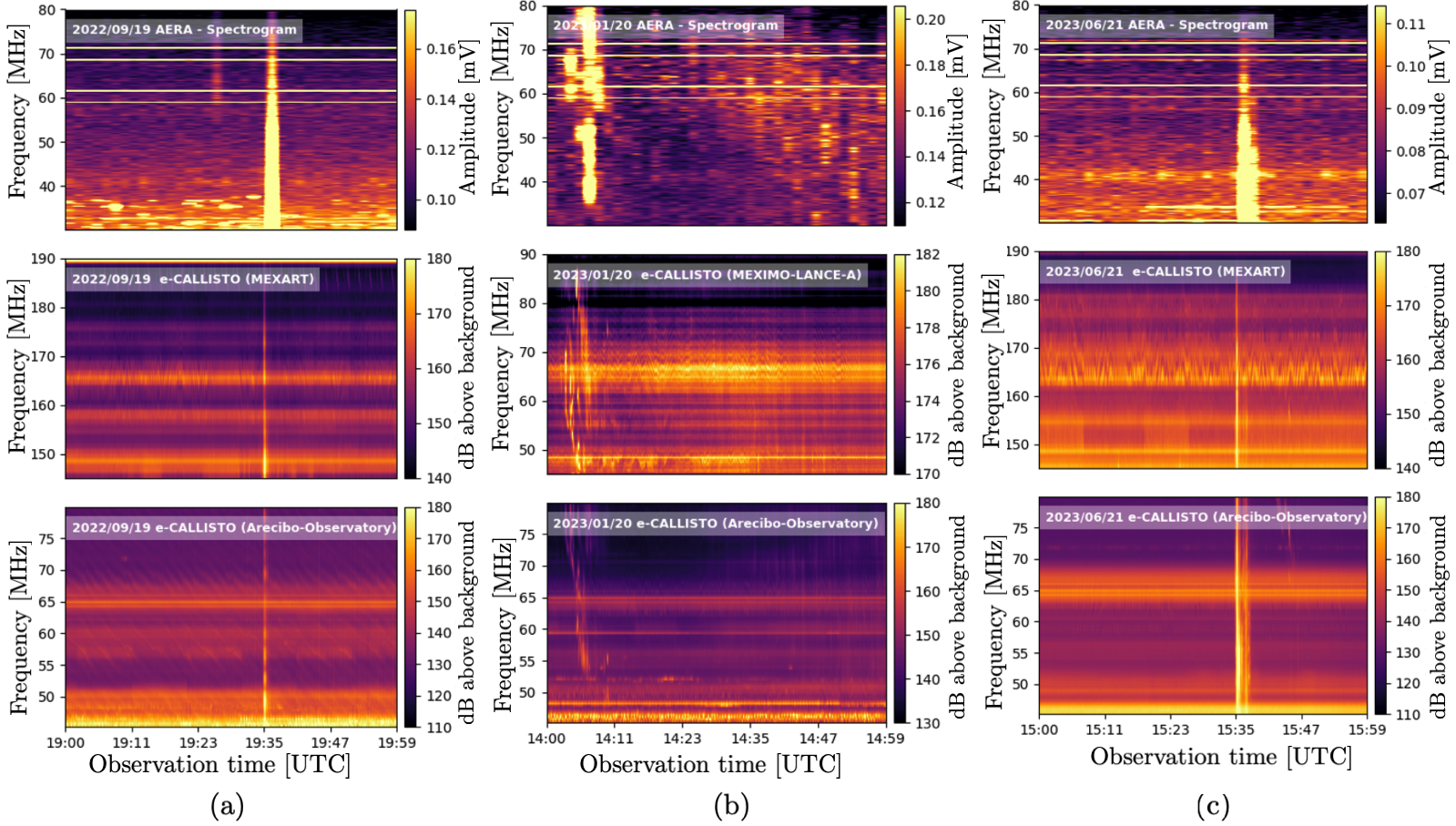}\quad
         \caption{\footnotesize{Spectrograms illustrating three examples of temporal coincidences between SRBs detected by AERA (top panel in each figure) and two different e-CALLISTO antennas (middle and bottom panels) located at distinct sites. Figure (a) corresponds to a Type V SRB, while figures (b) and (c) display Type II SRBs. In all cases, a clear temporal alignment of the radio signals is observed, highlighting the simultaneous detection by AERA and the e-CALLISTO network.}}
         \label{Fig:CoincidenciasCallisto}
         \vspace{-0.4cm}
 \end{figure}
 
 \begin{figure}[H]
        \centering
        \vspace{-0.3cm}
                \includegraphics[scale=0.35]{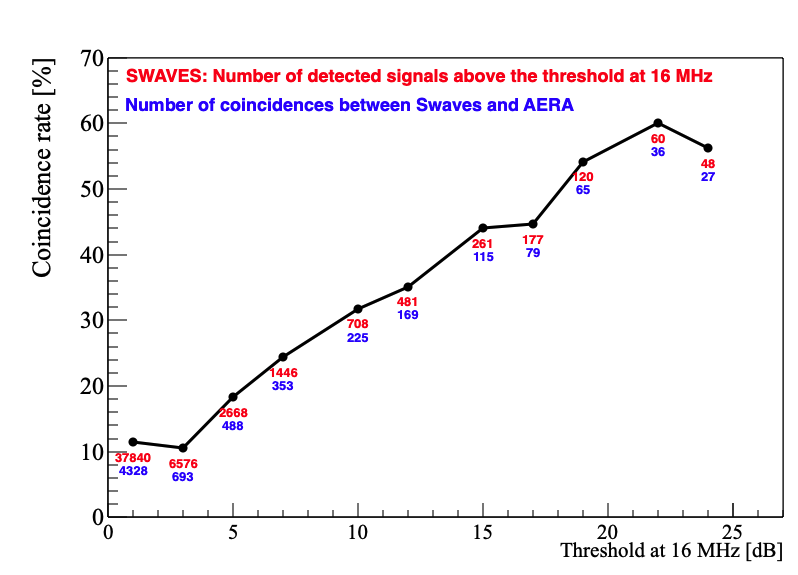}\quad
         \caption{\footnotesize{Temporal coincidence rate between the SRBs detected by AERA and those observed by SWAVES for different threshold levels in dB at its highest frequency (16 MHz). }}
         \label{Fig:CoincidenciasSWAVES}
         \vspace{-0.3cm}
 \end{figure}\hspace{-0.8cm}the algorithm applied to AERA data. Figure \ref{Fig:CoincidenciasCallisto} presents three examples of temporal coincidences observed between different antennas of the e-CALLISTO network and AERA stations.

We also performed a temporal coincidence search between solar events identified by the AERA detection algorithm and those observed by SWAVES during overlapping operational periods.  Although it operates in a different and lower frequency range compared to AERA’s terrestrial antennas, SWAVES reliably detects intense solar radio events. We focused on the 16 MHz band, which is the highest frequency measured by SWAVES and, therefore, the closest to AERA’s operational range. As shown in Figure \ref{Fig:CoincidenciasSWAVES}, the coincidence rate between SRBs detected by AERA and SWAVES is reported as a function of the threshold, in dB, used to filter SWAVES events. As expected, the coincidence rate increases with the applied threshold, reaching $\sim 50\%$ for events detected by SWAVES with signal strength above $\sim 20$ dB.

\section{Conclusions}
\label{conclusions}
We analyzed the impact of solar activity on AERA data collected over a decade. Our study identified a strong correlation between the MUF, modulated by the solar cycle, and broadband noise in the $30-40$ MHz range, indicating that this noise is likely caused by enhanced ionization in the upper ionosphere, leading to increased atmospheric reflection of terrestrial radio waves. 

Additionally, we searched for radio blackouts reported by NOAA between 2014 and 2024 that occurred during daylight hours in South America. Out of the 16 events associated with moderate to strong radio blackouts, in six showed clear signs of blackout in the AERA data. Interestingly, several of these events also exhibited signatures of solar radio bursts. We also searched for time coincidences between SRBs measured by AERA and independent measurements from the e-CALLISTO network and the SWAVES instrument. There were 204 coincidences with e-CALLISTO, and a coincidence rate of $\sim 50\%$ was achieved with SWAVES when selecting events detected by SWAVES with signal strength above $\sim 20$ dB at 16 MHz.
Although AERA was originally designed for detecting radio waves from extensive air showers, our findings reveal its broader potential for studying solar activity and its effects on Earth’s atmosphere. In addition, identifying periods affected by solar radio bursts and radio blackouts is essential, as such events may interfere with the quality of the recorded data and should be considered when selecting clean datasets for cosmic ray analyses. The Pierre Auger Observatory, with its integrated radio antennas and ground-based particle detectors, enables the detection of solar radio bursts, as demonstrated in this work, and also has the potential to measure the arrival of charged particles associated with these events using the particle detectors.

\newpage

\section*{The Pierre Auger Collaboration}

{\footnotesize\setlength{\baselineskip}{10pt}
\noindent
\begin{wrapfigure}[11]{l}{0.12\linewidth}
\vspace{-4pt}
\includegraphics[width=0.98\linewidth]{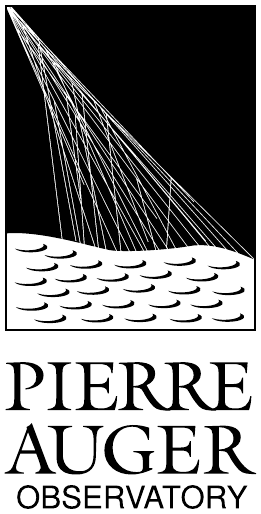}
\end{wrapfigure}
\begin{sloppypar}\noindent
A.~Abdul Halim$^{13}$,
P.~Abreu$^{70}$,
M.~Aglietta$^{53,51}$,
I.~Allekotte$^{1}$,
K.~Almeida Cheminant$^{78,77}$,
A.~Almela$^{7,12}$,
R.~Aloisio$^{44,45}$,
J.~Alvarez-Mu\~niz$^{76}$,
A.~Ambrosone$^{44}$,
J.~Ammerman Yebra$^{76}$
% ...

\end{sloppypar}
\begin{center}
\end{center}

\vspace{1ex}
\begin{description}[labelsep=0.2em,align=right,labelwidth=0.7em,labelindent=0em,leftmargin=2em,noitemsep,before={\renewcommand\makelabel[1]{##1 }}]
\item[$^{1}$] Centro At\'omico Bariloche and Instituto Balseiro (CNEA-UNCuyo-CONICET), San Carlos de Bariloche, Argentina
\item[$^{2}$] Departamento de F\'\i{}sica and Departamento de Ciencias de la Atm\'osfera y los Oc\'eanos, FCEyN, Universidad de Buenos Aires and CONICET, Buenos Aires, Argentina
\item[$^{3}$] IFLP, Universidad Nacional de La Plata and CONICET, La Plata, Argentina
\item[$^{4}$] Instituto de Astronom\'\i{}a y F\'\i{}sica del Espacio (IAFE, CONICET-UBA), Buenos Aires, Argentina
\item[$^{5}$] Instituto de F\'\i{}sica de Rosario (IFIR) -- CONICET/U.N.R.\ and Facultad de Ciencias Bioqu\'\i{}micas y Farmac\'euticas U.N.R., Rosario, Argentina
\item[$^{6}$] Instituto de Tecnolog\'\i{}as en Detecci\'on y Astropart\'\i{}culas (CNEA, CONICET, UNSAM), and Universidad Tecnol\'ogica Nacional -- Facultad Regional Mendoza (CONICET/CNEA), Mendoza, Argentina
\item[$^{7}$] Instituto de Tecnolog\'\i{}as en Detecci\'on y Astropart\'\i{}culas (CNEA, CONICET, UNSAM), Buenos Aires, Argentina
\item[$^{8}$] International Center of Advanced Studies and Instituto de Ciencias F\'\i{}sicas, ECyT-UNSAM and CONICET, Campus Miguelete -- San Mart\'\i{}n, Buenos Aires, Argentina
\item[$^{9}$] Laboratorio Atm\'osfera -- Departamento de Investigaciones en L\'aseres y sus Aplicaciones -- UNIDEF (CITEDEF-CONICET), Argentina
\item[$^{10}$] Observatorio Pierre Auger, Malarg\"ue, Argentina
\item[$^{11}$] Observatorio Pierre Auger and Comisi\'on Nacional de Energ\'\i{}a At\'omica, Malarg\"ue, Argentina
\item[$^{12}$] Universidad Tecnol\'ogica Nacional -- Facultad Regional Buenos Aires, Buenos Aires, Argentina
\item[$^{13}$] University of Adelaide, Adelaide, S.A., Australia
\item[$^{14}$] Universit\'e Libre de Bruxelles (ULB), Brussels, Belgium
\item[$^{15}$] Vrije Universiteit Brussels, Brussels, Belgium
\item[$^{16}$] Centro Brasileiro de Pesquisas Fisicas, Rio de Janeiro, RJ, Brazil
\item[$^{17}$] Centro Federal de Educa\c{c}\~ao Tecnol\'ogica Celso Suckow da Fonseca, Petropolis, Brazil
\item[$^{18}$] Instituto Federal de Educa\c{c}\~ao, Ci\^encia e Tecnologia do Rio de Janeiro (IFRJ), Brazil
\item[$^{19}$] Universidade de S\~ao Paulo, Escola de Engenharia de Lorena, Lorena, SP, Brazil
\item[$^{20}$] Universidade de S\~ao Paulo, Instituto de F\'\i{}sica de S\~ao Carlos, S\~ao Carlos, SP, Brazil
\item[$^{21}$] Universidade de S\~ao Paulo, Instituto de F\'\i{}sica, S\~ao Paulo, SP, Brazil
\item[$^{22}$] Universidade Estadual de Campinas (UNICAMP), IFGW, Campinas, SP, Brazil
\item[$^{23}$] Universidade Estadual de Feira de Santana, Feira de Santana, Brazil
\item[$^{24}$] Universidade Federal de Campina Grande, Centro de Ciencias e Tecnologia, Campina Grande, Brazil
\item[$^{25}$] Universidade Federal do ABC, Santo Andr\'e, SP, Brazil
\item[$^{26}$] Universidade Federal do Paran\'a, Setor Palotina, Palotina, Brazil
\item[$^{27}$] Universidade Federal do Rio de Janeiro, Instituto de F\'\i{}sica, Rio de Janeiro, RJ, Brazil
\item[$^{28}$] Universidad de Medell\'\i{}n, Medell\'\i{}n, Colombia
\item[$^{29}$] Universidad Industrial de Santander, Bucaramanga, Colombia
\item[$^{30}$] Charles University, Faculty of Mathematics and Physics, Institute of Particle and Nuclear Physics, Prague, Czech Republic
\item[$^{31}$] Institute of Physics of the Czech Academy of Sciences, Prague, Czech Republic
\item[$^{32}$] Palacky University, Olomouc, Czech Republic
\item[$^{33}$] CNRS/IN2P3, IJCLab, Universit\'e Paris-Saclay, Orsay, France
\item[$^{34}$] Laboratoire de Physique Nucl\'eaire et de Hautes Energies (LPNHE), Sorbonne Universit\'e, Universit\'e de Paris, CNRS-IN2P3, Paris, France
\item[$^{35}$] Univ.\ Grenoble Alpes, CNRS, Grenoble Institute of Engineering Univ.\ Grenoble Alpes, LPSC-IN2P3, 38000 Grenoble, France
\item[$^{36}$] Universit\'e Paris-Saclay, CNRS/IN2P3, IJCLab, Orsay, France
\item[$^{37}$] Bergische Universit\"at Wuppertal, Department of Physics, Wuppertal, Germany
\item[$^{38}$] Karlsruhe Institute of Technology (KIT), Institute for Experimental Particle Physics, Karlsruhe, Germany
\item[$^{39}$] Karlsruhe Institute of Technology (KIT), Institut f\"ur Prozessdatenverarbeitung und Elektronik, Karlsruhe, Germany
\item[$^{40}$] Karlsruhe Institute of Technology (KIT), Institute for Astroparticle Physics, Karlsruhe, Germany
\item[$^{41}$] RWTH Aachen University, III.\ Physikalisches Institut A, Aachen, Germany
\item[$^{42}$] Universit\"at Hamburg, II.\ Institut f\"ur Theoretische Physik, Hamburg, Germany
\item[$^{43}$] Universit\"at Siegen, Department Physik -- Experimentelle Teilchenphysik, Siegen, Germany
\item[$^{44}$] Gran Sasso Science Institute, L'Aquila, Italy
\item[$^{45}$] INFN Laboratori Nazionali del Gran Sasso, Assergi (L'Aquila), Italy
\item[$^{46}$] INFN, Sezione di Catania, Catania, Italy
\item[$^{47}$] INFN, Sezione di Lecce, Lecce, Italy
\item[$^{48}$] INFN, Sezione di Milano, Milano, Italy
\item[$^{49}$] INFN, Sezione di Napoli, Napoli, Italy
\item[$^{50}$] INFN, Sezione di Roma ``Tor Vergata'', Roma, Italy
\item[$^{51}$] INFN, Sezione di Torino, Torino, Italy
\item[$^{52}$] Istituto di Astrofisica Spaziale e Fisica Cosmica di Palermo (INAF), Palermo, Italy
\item[$^{53}$] Osservatorio Astrofisico di Torino (INAF), Torino, Italy
\item[$^{54}$] Politecnico di Milano, Dipartimento di Scienze e Tecnologie Aerospaziali , Milano, Italy
\item[$^{55}$] Universit\`a del Salento, Dipartimento di Matematica e Fisica ``E.\ De Giorgi'', Lecce, Italy
\item[$^{56}$] Universit\`a dell'Aquila, Dipartimento di Scienze Fisiche e Chimiche, L'Aquila, Italy
\item[$^{57}$] Universit\`a di Catania, Dipartimento di Fisica e Astronomia ``Ettore Majorana``, Catania, Italy
\item[$^{58}$] Universit\`a di Milano, Dipartimento di Fisica, Milano, Italy
\item[$^{59}$] Universit\`a di Napoli ``Federico II'', Dipartimento di Fisica ``Ettore Pancini'', Napoli, Italy
\item[$^{60}$] Universit\`a di Palermo, Dipartimento di Fisica e Chimica ''E.\ Segr\`e'', Palermo, Italy
\item[$^{61}$] Universit\`a di Roma ``Tor Vergata'', Dipartimento di Fisica, Roma, Italy
\item[$^{62}$] Universit\`a Torino, Dipartimento di Fisica, Torino, Italy
\item[$^{63}$] Benem\'erita Universidad Aut\'onoma de Puebla, Puebla, M\'exico
\item[$^{64}$] Unidad Profesional Interdisciplinaria en Ingenier\'\i{}a y Tecnolog\'\i{}as Avanzadas del Instituto Polit\'ecnico Nacional (UPIITA-IPN), M\'exico, D.F., M\'exico
\item[$^{65}$] Universidad Aut\'onoma de Chiapas, Tuxtla Guti\'errez, Chiapas, M\'exico
\item[$^{66}$] Universidad Michoacana de San Nicol\'as de Hidalgo, Morelia, Michoac\'an, M\'exico
\item[$^{67}$] Universidad Nacional Aut\'onoma de M\'exico, M\'exico, D.F., M\'exico
\item[$^{68}$] Institute of Nuclear Physics PAN, Krakow, Poland
\item[$^{69}$] University of \L{}\'od\'z, Faculty of High-Energy Astrophysics,\L{}\'od\'z, Poland
\item[$^{70}$] Laborat\'orio de Instrumenta\c{c}\~ao e F\'\i{}sica Experimental de Part\'\i{}culas -- LIP and Instituto Superior T\'ecnico -- IST, Universidade de Lisboa -- UL, Lisboa, Portugal
\item[$^{71}$] ``Horia Hulubei'' National Institute for Physics and Nuclear Engineering, Bucharest-Magurele, Romania
\item[$^{72}$] Institute of Space Science, Bucharest-Magurele, Romania
\item[$^{73}$] Center for Astrophysics and Cosmology (CAC), University of Nova Gorica, Nova Gorica, Slovenia
\item[$^{74}$] Experimental Particle Physics Department, J.\ Stefan Institute, Ljubljana, Slovenia
\item[$^{75}$] Universidad de Granada and C.A.F.P.E., Granada, Spain
\item[$^{76}$] Instituto Galego de F\'\i{}sica de Altas Enerx\'\i{}as (IGFAE), Universidade de Santiago de Compostela, Santiago de Compostela, Spain
\item[$^{77}$] IMAPP, Radboud University Nijmegen, Nijmegen, The Netherlands
\item[$^{78}$] Nationaal Instituut voor Kernfysica en Hoge Energie Fysica (NIKHEF), Science Park, Amsterdam, The Netherlands
\item[$^{79}$] Stichting Astronomisch Onderzoek in Nederland (ASTRON), Dwingeloo, The Netherlands
\item[$^{80}$] Universiteit van Amsterdam, Faculty of Science, Amsterdam, The Netherlands
\item[$^{81}$] Case Western Reserve University, Cleveland, OH, USA
\item[$^{82}$] Colorado School of Mines, Golden, CO, USA
\item[$^{83}$] Department of Physics and Astronomy, Lehman College, City University of New York, Bronx, NY, USA
\item[$^{84}$] Michigan Technological University, Houghton, MI, USA
\item[$^{85}$] New York University, New York, NY, USA
\item[$^{86}$] University of Chicago, Enrico Fermi Institute, Chicago, IL, USA
\item[$^{87}$] University of Delaware, Department of Physics and Astronomy, Bartol Research Institute, Newark, DE, USA
\item[] -----
\item[$^{a}$] Max-Planck-Institut f\"ur Radioastronomie, Bonn, Germany
\item[$^{b}$] also at Kapteyn Institute, University of Groningen, Groningen, The Netherlands
\item[$^{c}$] School of Physics and Astronomy, University of Leeds, Leeds, United Kingdom
\item[$^{d}$] Fermi National Accelerator Laboratory, Fermilab, Batavia, IL, USA
\item[$^{e}$] Pennsylvania State University, University Park, PA, USA
\item[$^{f}$] Colorado State University, Fort Collins, CO, USA
\item[$^{g}$] Louisiana State University, Baton Rouge, LA, USA
\item[$^{h}$] now at Graduate School of Science, Osaka Metropolitan University, Osaka, Japan
\item[$^{i}$] Institut universitaire de France (IUF), France
\item[$^{j}$] now at Technische Universit\"at Dortmund and Ruhr-Universit\"at Bochum, Dortmund and Bochum, Germany
\end{description}

\section*{Acknowledgments}

\begin{sloppypar}
The successful installation, commissioning, and operation of the Pierre
Auger Observatory would not have been possible without the strong
commitment and effort from the technical and administrative staff in
Malarg\"ue. We are very grateful to the following agencies and
organizations for financial support:
\end{sloppypar}

\begin{sloppypar}
Argentina -- Comisi\'on Nacional de Energ\'\i{}a At\'omica; Agencia Nacional de
Promoci\'on Cient\'\i{}fica y Tecnol\'ogica (ANPCyT); Consejo Nacional de
Investigaciones Cient\'\i{}ficas y T\'ecnicas (CONICET); Gobierno de la
Provincia de Mendoza; Municipalidad de Malarg\"ue; NDM Holdings and Valle
Las Le\~nas; in gratitude for their continuing cooperation over land
access; Australia -- the Australian Research Council; Belgium -- Fonds
de la Recherche Scientifique (FNRS); Research Foundation Flanders (FWO),
Marie Curie Action of the European Union Grant No.~101107047; Brazil --
Conselho Nacional de Desenvolvimento Cient\'\i{}fico e Tecnol\'ogico (CNPq);
Financiadora de Estudos e Projetos (FINEP); Funda\c{c}\~ao de Amparo \`a
Pesquisa do Estado de Rio de Janeiro (FAPERJ); S\~ao Paulo Research
Foundation (FAPESP) Grants No.~2019/10151-2, No.~2010/07359-6 and
No.~1999/05404-3; Minist\'erio da Ci\^encia, Tecnologia, Inova\c{c}\~oes e
Comunica\c{c}\~oes (MCTIC); Czech Republic -- GACR 24-13049S, CAS LQ100102401,
MEYS LM2023032, CZ.02.1.01/0.0/0.0/16{\textunderscore}013/0001402,
CZ.02.1.01/0.0/0.0/18{\textunderscore}046/0016010 and
CZ.02.1.01/0.0/0.0/17{\textunderscore}049/0008422 and CZ.02.01.01/00/22{\textunderscore}008/0004632;
France -- Centre de Calcul IN2P3/CNRS; Centre National de la Recherche
Scientifique (CNRS); Conseil R\'egional Ile-de-France; D\'epartement
Physique Nucl\'eaire et Corpusculaire (PNC-IN2P3/CNRS); D\'epartement
Sciences de l'Univers (SDU-INSU/CNRS); Institut Lagrange de Paris (ILP)
Grant No.~LABEX ANR-10-LABX-63 within the Investissements d'Avenir
Programme Grant No.~ANR-11-IDEX-0004-02; Germany -- Bundesministerium
f\"ur Bildung und Forschung (BMBF); Deutsche Forschungsgemeinschaft (DFG);
Finanzministerium Baden-W\"urttemberg; Helmholtz Alliance for
Astroparticle Physics (HAP); Helmholtz-Gemeinschaft Deutscher
Forschungszentren (HGF); Ministerium f\"ur Kultur und Wissenschaft des
Landes Nordrhein-Westfalen; Ministerium f\"ur Wissenschaft, Forschung und
Kunst des Landes Baden-W\"urttemberg; Italy -- Istituto Nazionale di
Fisica Nucleare (INFN); Istituto Nazionale di Astrofisica (INAF);
Ministero dell'Universit\`a e della Ricerca (MUR); CETEMPS Center of
Excellence; Ministero degli Affari Esteri (MAE), ICSC Centro Nazionale
di Ricerca in High Performance Computing, Big Data and Quantum
Computing, funded by European Union NextGenerationEU, reference code
CN{\textunderscore}00000013; M\'exico -- Consejo Nacional de Ciencia y Tecnolog\'\i{}a
(CONACYT) No.~167733; Universidad Nacional Aut\'onoma de M\'exico (UNAM);
PAPIIT DGAPA-UNAM; The Netherlands -- Ministry of Education, Culture and
Science; Netherlands Organisation for Scientific Research (NWO); Dutch
national e-infrastructure with the support of SURF Cooperative; Poland
-- Ministry of Education and Science, grants No.~DIR/WK/2018/11 and
2022/WK/12; National Science Centre, grants No.~2016/22/M/ST9/00198,
2016/23/B/ST9/01635, 2020/39/B/ST9/01398, and 2022/45/B/ST9/02163;
Portugal -- Portuguese national funds and FEDER funds within Programa
Operacional Factores de Competitividade through Funda\c{c}\~ao para a Ci\^encia
e a Tecnologia (COMPETE); Romania -- Ministry of Research, Innovation
and Digitization, CNCS-UEFISCDI, contract no.~30N/2023 under Romanian
National Core Program LAPLAS VII, grant no.~PN 23 21 01 02 and project
number PN-III-P1-1.1-TE-2021-0924/TE57/2022, within PNCDI III; Slovenia
-- Slovenian Research Agency, grants P1-0031, P1-0385, I0-0033, N1-0111;
Spain -- Ministerio de Ciencia e Innovaci\'on/Agencia Estatal de
Investigaci\'on (PID2019-105544GB-I00, PID2022-140510NB-I00 and
RYC2019-027017-I), Xunta de Galicia (CIGUS Network of Research Centers,
Consolidaci\'on 2021 GRC GI-2033, ED431C-2021/22 and ED431F-2022/15),
Junta de Andaluc\'\i{}a (SOMM17/6104/UGR and P18-FR-4314), and the European
Union (Marie Sklodowska-Curie 101065027 and ERDF); USA -- Department of
Energy, Contracts No.~DE-AC02-07CH11359, No.~DE-FR02-04ER41300,
No.~DE-FG02-99ER41107 and No.~DE-SC0011689; National Science Foundation,
Grant No.~0450696, and NSF-2013199; The Grainger Foundation; Marie
Curie-IRSES/EPLANET; European Particle Physics Latin American Network;
and UNESCO.
\end{sloppypar}

}


\begin{thebibliography}{99}
\small
\setlength{\itemsep}{2pt}


\bibitem{AERA}P.Abreu \textit{et al}., JINST \textbf{7}  (2012) 10011

\bibitem{TypeI}Trakhtengerts, V. Yu., Soviet Astronomy, \textbf{10} (1966) 281

\bibitem{KUNDU}Kundu, Mukul R., New York: Interscience Publication (1965)

\bibitem{TypeII}Wijesekera, J. V., Journal of Physics: Conference Series  (2018) 012046

\bibitem{TypeIII}Reid, Hamish AS. Frontiers in Astronomy and Space Sciences \textbf{7} (2020) 56

\bibitem{TypeIV}Weiss, A. A. Australian Journal of Physics \textbf{16} (1963) 526

\bibitem{TypeV}Sheridan, K. V., The Observatory \textbf{79} (1959) 51


\bibitem{IRI}BILITZA, Dieter \textit{et al}. Reviews of geophysics \textbf{60} (2022) 4

\bibitem{GOES_NOAA}Hill, S. M. \textit{et al}., Solar Physics \textbf{226} (2005) 255

\bibitem{NOAA_scale}Hanslmeier, Arnold. The Sun and Space Weather (2002) 193

\bibitem{RNO-G} AGARWAL, S. \textit{et al}. Astroparticle physics, \textbf{164} (2025) 103024.

\bibitem{Callisto} Christian Monstein, \textit{et al}., CALLISTO Solar Spectrogram FITS files Data set (2023)

\bibitem{Swaves}KAISER, Michael L. \textit{et al}., Stereo mission overview. IEEE, 2007. p. 1-8.


\end{thebibliography}
\end{document}